\documentclass[Physsubmission, Phys]{SciPost}
\hypersetup{
    colorlinks,
    linkcolor={red!50!black},
    citecolor={blue!50!black},
    urlcolor={blue!80!black}
}

\usepackage[bitstream-charter]{mathdesign}
\usepackage{slashed}
\DeclareSymbolFont{usualmathcal}{OMS}{cmsy}{m}{n}
\DeclareSymbolFontAlphabet{\mathcal}{usualmathcal}

\newcommand{\MSbar}{\overline{\mathrm{MS}}}

\begin{document}
%\begin{flushright}
%\small TTP21-038, P3H-21-073
%\end{flushright}

\begin{center}{\Large \textbf{
Higher-order corrections to the kinetic mass definition for the bottom and the charm quarks
}}\end{center}

% TODO: write the author list here. Use initials + surname format.
% Separate subsequent authors by a comma, omit comma at the end of the list.
% Mark the corresponding author with a superscript *.
\begin{center}
Matteo Fael\textsuperscript{$\star$}
\end{center}

\begin{center}
Institut f{\"u}r Theoretische Teilchenphysik, Karlsruhe
Institute of Technology (KIT),\\ 76128 Karlsruhe, Germany
\\
$\star$ matteo.fael@kit.edu
\end{center}

% For convenience during refereeing (optional),
% you can turn on line numbers by uncommenting the next line:
%\linenumbers

\definecolor{palegray}{gray}{0.95}
\begin{center}
\colorbox{palegray}{
  \begin{tabular}{rr}
  \begin{minipage}{0.1\textwidth}
    \includegraphics[width=35mm]{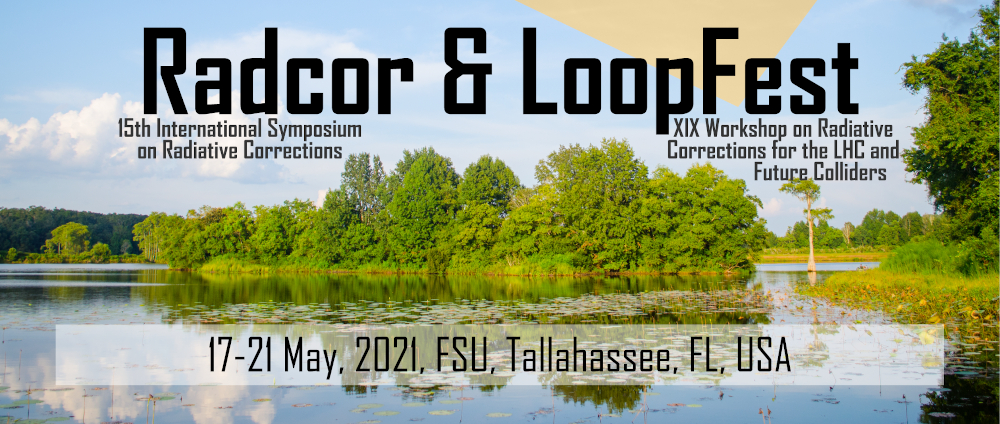}
  \end{minipage}
  &
  \begin{minipage}{0.85\textwidth}
    \begin{center}
    {\it 15th International Symposium on Radiative Corrections: \\Applications of Quantum Field Theory to Phenomenology,}\\
    {\it FSU, Tallahasse, FL, USA, 17-21 May 2021} \\
    %\doi{10.21468/SciPostPhysProc.?}\\
    \end{center}
  \end{minipage}
\end{tabular}
}
\end{center}

\section*{Abstract}
{\bf \boldmath
%Short distance masses are crucial ingredients in the prediction of inclusive semileptonic $B$-meson decays. 
In these proceedings we discuss the relation between the kinetic and the on-shell schemes for the bottom and the charm quarks and present the methods for the calculation of the mass relation to higher orders in perturbative QCD.
The bottom mass in the kinetic scheme is a pivotal input parameter in the inclusive determination of $|V_{cb}|$ from $B\to X_c \ell \nu_\ell$ decays.
By combining the relation between the kinetic and the on-shell mass with well-know results for the $\MSbar$-on-shell conversion, we obtain a prediction for $m_b^\mathrm{kin}$ based on precise determinations of $\overline{m}_b(\overline{m}_b)$. 
}

%\vspace{10pt}
%\noindent\rule{\textwidth}{1pt}
%\tableofcontents\thispagestyle{fancy}
%\noindent\rule{\textwidth}{1pt}
%\vspace{10pt}

\section{Introduction}
\label{sec:intro}
% TODO: write your article here.
The prediction of inclusive $B \to X_{c} \ell \nu_\ell$ decays is closely intertwined with the mass schemes for the charm and the bottom quarks. 
The theoretical framework  for this class of decays, the Heavy Quark Expansion (HQE), is well established and allows us to predict the semileptonic total rate and the moments of various kinematic spectra as a double expansion in the strong coupling constant $\alpha_s(m_b)$ and $\Lambda_{QCD}/m_b$.
The size of the parturbative QCD corrections strongly depends on the heavy-quark mass scheme, which is crucial to obtain precise predictions. The on-shell scheme is affected by renormalon ambiguity and yields a bad behaviour of the perturbative series~\cite{Beneke:1994sw,Bigi:1994em}. 
For instance, to leading order in $\Lambda_{QCD}/m_b$, the $\alpha_s$ expansion for the semileptonic width $\Gamma_\mathrm{sl}$ behaves as:
\begin{equation}
\Gamma_\mathrm{sl} = 
\frac{G_F^2 |V_{cb}|^2 (m_b^\mathrm{OS})^5}{192 \pi^3} f(\rho) 
\left[
1- 1.78 \left( \frac{\alpha_s}{\pi} \right)
-13.1 \left( \frac{\alpha_s}{\pi} \right)^2
-163.3 \left( \frac{\alpha_s}{\pi} \right)^3
\right]
+O\left( \frac{1}{m_b^2}\right),
\end{equation}
where $G_F$ is the Fermi constant, $f(\rho) =1-8\rho -12 \rho^2 \log \rho +8 \rho^3 -\rho^4 $, $\alpha_s \equiv \alpha_s^{(5)}(m_b^\mathrm{OS})$ and $\sqrt{\rho} =m_c^\mathrm{OS}/m_b^\mathrm{OS} = 0.25$. Both at $O(\alpha_s^2)$ and $O(\alpha_s^3)$ one observes a shift of about $-6\%$ which shows that the preturbative expansion is badly behaved.
Also in the $\MSbar$ scheme, the $\alpha_s$ corrections to $\Gamma_\mathrm{sl}$ have a bad convergence. Indeed, removing the infra-red (IR) renormalons by using a short distance mass definition does not guarantee a fast convergent perturbative series. 
The overall factor $m_b^5$ generates power-enhanced terms when converting to another scheme, \textit{e.g.} $ m_b^\mathrm{OS} = \tilde m_b [1+ c_1 \alpha_s/\pi + c_2 (\alpha_s/\pi)^2]$:
\begin{equation}
    \Gamma_\mathrm{sl} =
    \frac{G_F^2 |V_{cb}|^2 \tilde m_b^5}{192 \pi^3} f(\rho) 
    \left[
    1+(-1.78+5 c_1) \left( \frac{\alpha_s}{\pi} \right)
    +(-13.1-8.9 c_1+10 c_1^2+5 c_2) \left( \frac{\alpha_s}{\pi} \right)^2
\right]   
\end{equation}
Terms like $ \binom{5}{k} c_1^k \alpha_s^k$ can spoil the first orders in the perturbative series.
The heavy-quark \emph{kinetic scheme} ($m^\mathrm{kin}_Q$) was introduced in~\cite{Bigi:1994ga,Bigi:1996si} to resum in the semileptonic rate such enhanced terms via a suitable short-distance definition. Global fits of inclusive semileptonic decays in~\cite{Gambino:2013rza,Alberti:2014yda,Gambino:2016jkc,Bordone:2021oof} have used the kinetic scheme for the definition of the bottom mass and the non-perturbative parameters.
An alternative approach is given by the $1S$ scheme developed in~\cite{Hoang:1998ng,Hoang:1998hm,Hoang:1999us} which has been used also for the extraction of the inclusive $|V_{cb}|$ in~\cite{Bauer:2004ve}.

The relation between the on-shell mass and the kinetic mass was computed  up to $O(\alpha_s^2)$ more than 20 years ago~\cite{Czarnecki:1997sz}.
The need for a further improvement up to $O(\alpha_s^2)$ is two fold. On the one hand, a prediction of the partonic rate up to $O(\alpha_s^3)$, necessary to reduce the theoretical uncertainty on the semileptonic rate, requires the mass conversion formula up to the same order~\cite{Fael:2020tow}.
On the other hand, global fits must take advantage of several external constraints to improve the extraction of $|V_{cb}|$. 
The semileptonic $B$ decays alone precisely determine only a linear combination of the heavy quark masses, approximately given by $m_b-0.8m_c$~\cite{Gambino:2013rza}.  
So, in order to break the degeneracy one must include external  constraint  for  the  bottom  (or the  charm) mass. These mass values are obtained, for instance, from lattice QCD~\cite{FlavourLatticeAveragingGroup:2019iem} or sum rules~\cite{Chetyrkin:2009fv} which are usually given in the $\MSbar$ scheme.
Before the calculation in~\cite{Fael:2020njb,Fael:2020iea}, the $O(\alpha_s^2)$ scheme-conversion uncertainty of $m_b^\mathrm{kin}$ was much larger than the error of $\overline{m}_b(\overline{m}_b)$. 

In these proceedings we review the definition of the kinetic mass of the heavy quark and provide an explicit derivation of the leading order result.
We discuss also the strategy employed for the calculation of the $O(\alpha_s^3)$ corrections recently presented in~\cite{Fael:2020njb,Fael:2020iea}. 

\section{The kinetic mass}
\label{sec:kinmass}

The definition of the kinetic mass is based on the relation between the masses of a heavy meson $M_H$ and the corresponding heavy quark $m_Q$, derived within the Heavy Quark Effective Theory (HQET):
\begin{equation}
 M_H = m_Q + \overline \Lambda
 +\frac{\mu_\pi^2-\frac{d_H}{3} \mu_G^2}{2m_Q}
 %+\frac{\rho_D^3+\frac{d_H}{3} \rho_{LS}^3}{4m_Q^2}
 +O\left(\frac{1}{m_Q^2}\right),
 %-\frac{1}{4m_Q^2}
 %\left[
 % \rho_{\pi \pi}+ \rho_S 
 %+\frac{d_H}{3} \left(\rho_A + \rho_{\pi G}\right)
 %\right],
 \label{eqn:massrel}
\end{equation}
where $d_H=3$ for pseudo-scalar mesons and $d_H=-1$ for vector mesons. The parameters 
$\mu_\pi^2,\mu_G^2$ are matrix elements of local operators in HQET
while 
%$\rho_{\pi\pi}, \rho_S, \rho_A$ and $\rho_{\pi G}$ are non-local ones~\cite{Bigi:1994ga}.
$\overline{\Lambda}$ is the heavy  meson’s  binding  energy in the $m_Q \to \infty$ limit.

From~\eqref{eqn:massrel} we obtain the relation between the on-shell and the kinetic mass by identifying $M_H \to m_Q^\mathrm{OS}$ -- they both are scale independent -- and  $m_Q \to m_Q^\mathrm{kin}$; the relevant HQET parameters are computed in perturbation theory.
The averaged meson mass provides the definition of the mass relation up to order $1/m_Q$~\cite{Bigi:1996si}:
\begin{equation}
  m_Q^\mathrm{OS} = m_Q^\mathrm{kin}(\mu)
  + [\overline \Lambda(\mu)]_\mathrm{pert}
 +\frac{[\mu_\pi^2(\mu)]_\mathrm{pert}}{2m_Q^\mathrm{kin}(\mu)}
 + O\left(\frac{1}{m_Q^2} \right).
\end{equation}
The scale $\mu$ entering the definition of $m_Q^\mathrm{kin}$ and the parturbative version of the HQET parameters plays the role of a \emph{Wilsonian cutoff} with $\Lambda_\mathrm{QCD} \ll \mu \ll m_Q$. It allows to separate short- and long-distance effects in the heavy quark decay.

\begin{figure}[htb]
    \centering
    \begin{minipage}[b]{0.3\textwidth}
     \centering
     \includegraphics[width=\textwidth]{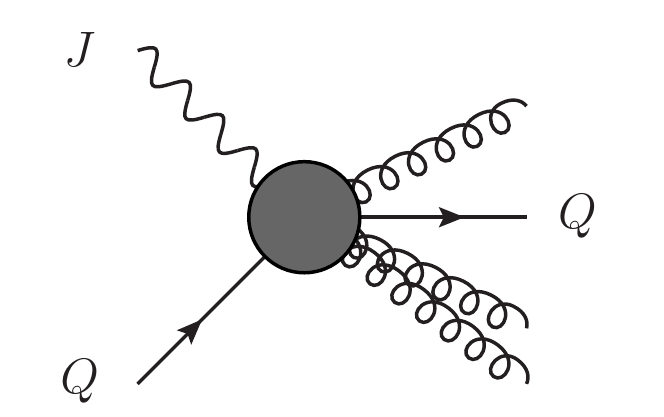}\\
     (a)
    \end{minipage}
    \begin{minipage}[b]{0.3\textwidth}
     \centering
     \includegraphics[width=0.85\textwidth]{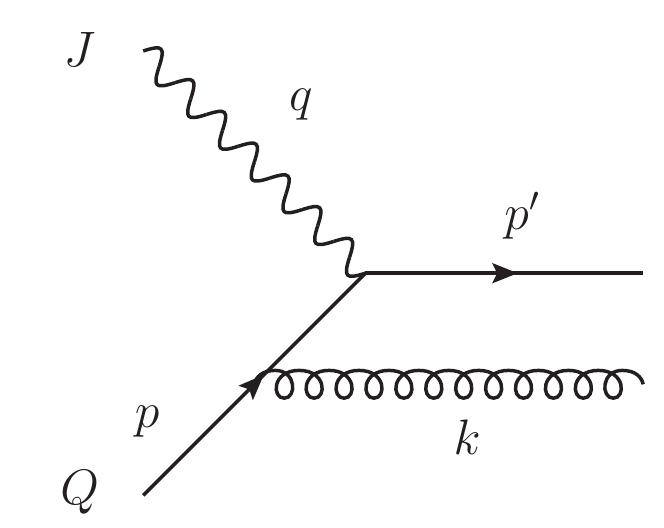}\\
     (b)
    \end{minipage}
    \begin{minipage}[b]{0.3\textwidth}
     \centering
     \includegraphics[width=0.85\textwidth]{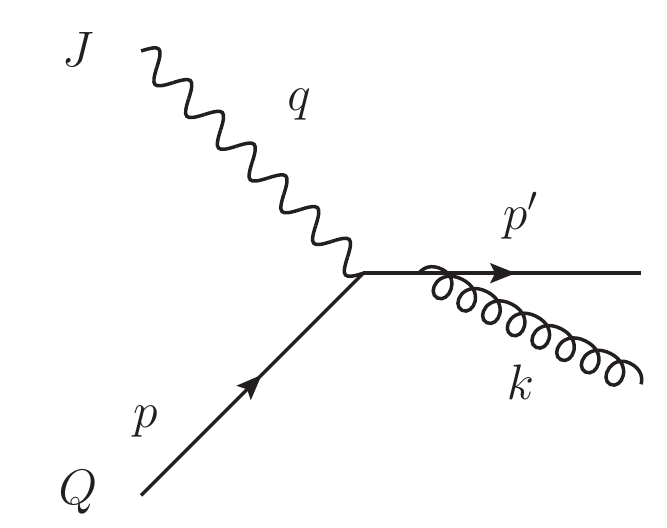}\\
     (c)
    \end{minipage}
    \caption{Scattering of a heavy quark $Q$ out of a current $J$ (left). 
    Feynman diagrams for the single-gluon emission at $O(\alpha_S)$ (center, right).}
    \label{fig:bJ}
\end{figure}
The perturbative versions of the HQET parameters are most conveniently computed by making use of the Small Velocity (SV) sum rules~\cite{Bigi:1994ga}. 
We consider the scattering of a heavy quark $Q$ induced by a generic current $J$ (as shown in Fig.~\ref{fig:bJ}(a)). The exact nature of the current $J$ is irrelevant, we can use for convenience a scalar current $J = \bar{Q}(x) Q(x)$ or a vector current $J^\mu =  \bar{Q}(x) \gamma^\mu Q(x)$. 
The current $J$ excites the quark $Q$ causing the emission of gluons and/or light quarks. We denote the generic multi-particle final state with $X_Q$ and we work in the rest frame of the initial quark $Q$.
The excitation energy $\omega$ of $X_Q$ is defined by
\begin{equation}
    \omega = q_0 - q_0^\mathrm{min} = q_0 - \left(\sqrt{\vec{q}^2+m_Q^2}-m_Q\right),
\end{equation}
where $q = (q_0,\vec q)$ is the four-momentum transferred by the current $J$ to the system.
Let us denote with $W(\omega,\vec v)$ the \emph{structure function} for the process, i.e.\ the squared matrix element, and with $\vec v = \vec q/m_Q$ the \emph{velocity} of the state $X_Q$.

We consider such scattering process in the heavy-quark mass limit $m_Q \to \infty$ and in the SV limit, i.e.\ the limit in which the $X_Q$'s velocity is small $|\vec v| \ll 1$. 
The Operator Product Expansion tells us that in this limit we can relate the moments of $W(\omega,\vec v)$ w.r.t.\ the excitation energy $\omega$ to the HQET parameters, in particular
\begin{align}
    [\overline{\Lambda}({\mu})]_\mathrm{pert} &= 
    \lim_{\vec v \to 0} \, \lim_{m_Q \to 0}\, 
    \frac{2}{\vec v \, ^2} \frac{
    \int_{0}^{{\mu}}
      \omega \,
    W(\omega,\vec v)
    \, d \omega
    }{
    \int_{0}^{\mu}
    \,
    W(\omega,\vec v)
    \, d \omega
    }\, ,
    \notag \\[15pt]
    [\mu_\pi^2 ({\mu})]_\mathrm{pert} &= 
    \lim_{\vec v \to 0} \, \lim_{m_Q \to 0}
    \frac{3}{\vec v \, ^2} \,
    \frac{
    \int_{0}^{{ \mu}}
     \omega^2 \,
    W(\omega,\vec v)
    \, d\omega 
    }{
    \int_{0}^{\mu}
     \,
    W(\omega,\vec v)
    \, d \omega
    }.
    \label{eqn:SV}
\end{align}
These are the SV sum rules which provide us with the rigorous definition of the perturbative HQET parameters in the kinetic mass formula. 

\section{The kinetic mass as a threshold mass}
\begin{figure}[htb]
    \centering
    \begin{minipage}[b]{0.3\textwidth}
     \includegraphics[width=\textwidth]{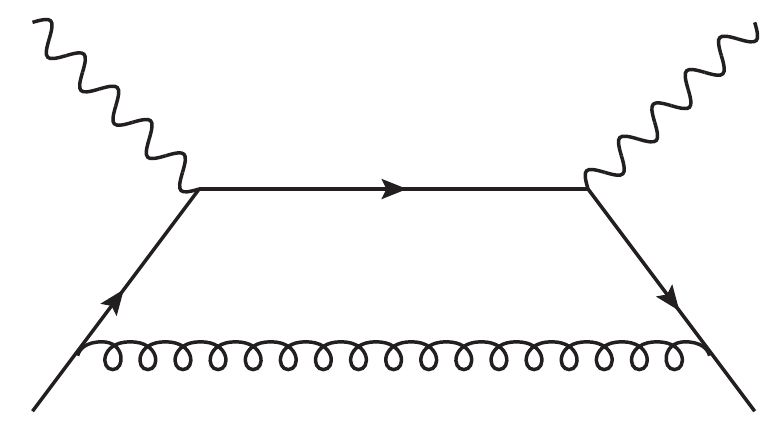}
    \end{minipage}
    \begin{minipage}[b]{0.3\textwidth}
     \includegraphics[width=\textwidth]{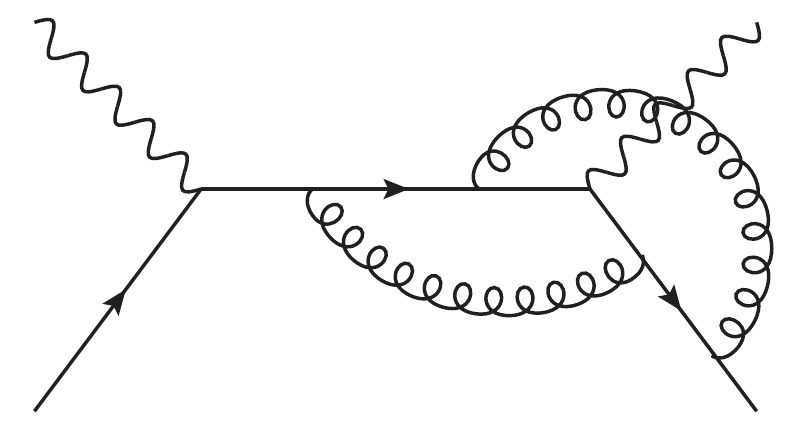}
    \end{minipage}
    \begin{minipage}[b]{0.3\textwidth}
     \includegraphics[width=\textwidth]{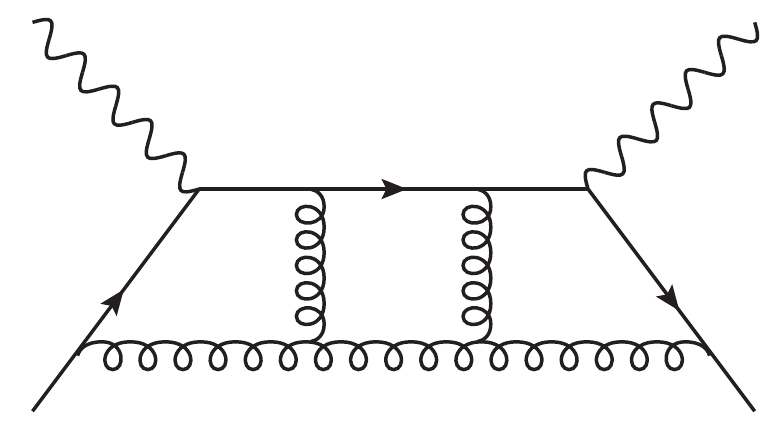}
    \end{minipage}
    \caption{Sample  Feynman  diagrams  for  the  scattering  process  of  an  external  current (wavy line) and a heavy quark (solid line).  Gluons are represented by curly lines.}
    \label{fig:scat}
\end{figure}

To calculate the relation between the kinetic mass and the on-shell mass in (2) we have to consider all possible perturbative QCD corrections to the structure function $W(\omega,\vec v)$, order by order in $\alpha_s$.
To this end in~\cite{Fael:2020iea,Fael:2020njb} we developed the following strategy.
\begin{itemize}
    \item We use the optical theorem to compute the structure function $W$, i.e.\ we
    consider multi-loop diagrams for $bJ \to bJ$ forward scattering (see Fig.~\ref{fig:scat}) and take their discontinuity.
    \item We calculate $W$ only to leading order in a $\vec v^2$ and  $\omega$ expansion. 
    This means we are interested in the following decomposition
    \begin{equation}
        W(\omega,\vec v) = 
        W_\mathrm{el} \delta(\omega) 
        + W_\mathrm{real} \frac{\vec v^2}{\omega } 
        \theta(\omega) 
        + O(\vec v^4, \omega^0).
        \label{eq::W}
    \end{equation}
    The first term on the r.h.s.\ arises from tree-level and all-virtual diagrams (which have zero excitation energy) while $W_\mathrm{real}$ comes from real emission contributions.
    \item Insert the $\alpha_s$ expansion of $W_\mathrm{el}$ and  $W_\mathrm{real}$ into the SV sum rules and re-expand the ratios in $\alpha_s$. 
    For instance for $\mu_\pi$ we have:
\begin{equation}
  [\mu_\pi(\mu)]_{\rm pert} =
      \lim_{\vec{v}\to0}\lim_{m_Q\to\infty} 
      \frac{3}{\vec{v}\,^2} 
      %\frac{\displaystyle
        \sum_{n=1} \alpha_s^n
	\int_0^\mu \omega^2 \,
         \frac{\vec{v}\,^2}{\omega}
       W_{\rm real}^{(n)}(\omega,\vec v) \, {\rm d}\omega
      %}
      %{
      %\displaystyle
      \Bigg/
      \sum_{n=0} \alpha_s^n 
      %\left(
      W_{\rm el}^{(n)}
      %+
      %\int_0^\mu \frac{\vec{v}\, ^2}{\omega} 
      %W_{\rm real}^{(n)} (\omega,\vec{v}\,) \, {\rm d}\omega
    %\right)
    %} 
    \,.
    \label{eq::Lam_2}
\end{equation}
Equation (\ref{eq::Lam_2}) shows that terms of $O(\vec v^4)$ or higher are eliminated by the limit $\vec{v}\to 0$.  
Moreover, we retain only the leading $1/\omega$ part since higher orders, which scale as $(\omega/m_Q)^n$, are eliminated by the limit $m_Q\to + \infty$.  
Due to the factors $\omega^k$ in the integrand of the numerator,
the $\delta$-function distribution in Eq.~(\ref{eq::W}) does not contribute to the numerator but only to the denominator. Thus virtual corrections are needed to one order less than the real ones. 
Vice versa, we discard real corrections at the denominator since, after expanding in $\alpha_s$, they become of order $\vec v \,^4$ and
vanish in the $\vec v \to 0$ limit.

\item For the practical calculation, we express the non-relativistic
quantities $\omega$ and $\vec{v}$ in terms of Lorenz invariants.
Let us define
\begin{eqnarray}
  y &\equiv& m_Q^2 - s = - \omega \left( 2m_Q \sqrt{ 1+\vec{v}\,^2 } + \omega \right)
  = - m_Q \, \omega ( 2 + \vec{v}\,^2 ) + {O}(\omega^2,\vec{v}\,^4 )
  \label{eq:scaling1}
        \,, \\
  q^2 &\equiv& \left[ m_Q \left( \sqrt{ 1+\vec{v}\,^2 } - 1 \right) + \omega \right]^2 - m_Q^2 \vec{v}\,^2
  =  - m_Q \, \vec{v}\,^2  ( m_Q - \omega ) + {O}(\omega^2,\vec{v}\,^4 )
  \label{eq:scaling2}
  \,.
\end{eqnarray}
This allows us to take the non-relativistic limits $\lim_{\vec{v}\to0}$ and $\lim_{m_Q\to\infty}$ by expanding $W$ around the \emph{one-particle threshold limit}  at $s=(p+q)^2=m_Q^2$ and subsequently expand in $q$.
In fact, the $\lim_{m_Q\to\infty}$ limit corresponds to an expansion in
 $ y = m_Q^2-s \le 0 $,
the difference between the Mandelstam variable $s$ and the position of the threshold at $s=m_Q^2$.
We realise the expansion with the help of the {\it method of regions}~\cite{Beneke:1997zp,Smirnov:2012gma}.  
The expansion $\vec{v}\to0$, on the other hand, reduces to a naive Taylor expansion in $q$. 
\end{itemize}

As an explicit example, let us derive the mass relation to order $\alpha_s$.
We start with the structure function at tree-level: $W^{(0)}$.
The scattering amplitude is simply
$
    i \mathcal M^{(0)} = 
    \overline u (p') \Gamma u(p) 
$
at $O(\alpha_s^0)$, where $p$ ($p'$) is the momentum of the incoming (outgoing) heavy quark.
We do not need to specify the current $J = \overline{Q}(x) \Gamma Q(x)$, 
as its dependence  will drop out in the end.
$W^{(0)}$ is obtained by performing the one-particle phase-space integration:
\begin{equation}
    W^{(0)} = \int \frac{d^3 p'}{(2\pi)^3 2p'_0} 
    |\mathcal M^{(0)}|^2 (2\pi)^4 \delta^4 (p+q-p')=
    (2\pi) \delta(s-m_Q^2) |\mathcal M^{(0)}|^2,
\end{equation}
as expected by the optical theorem. The integration w.r.t.\ $\omega$ gives
\begin{equation}
    \int_0^\mu \mathrm{d}\omega \, (2\pi) \delta(s-m_Q^2) |\mathcal M^{(0)}|^2 =
    \int_0^\mu \mathrm{d}\omega (2\pi) \delta(-y ) |\mathcal M^{(0)}|^2 
    %=\frac{\pi}{m_Q \sqrt{1+\vec v^2}} |\mathcal M^{(0)}|^2
    \simeq \frac{\pi}{m_Q} |\mathcal M^{(0)}|^2,
\end{equation}
since $y \sim - 2 m_Q \omega$. This is the denominator in Eq.~\eqref{eq::Lam_2}.
We define for later discussion
\begin{align}
    H &= \frac{\pi}{m_Q} |\mathcal M^{(0)}|^2, &
    U^{(0)} (\omega,\vec v)&= 2 \delta(s-m_Q^2) = 2 \delta(-y).
\end{align}

We then consider the amplitudes for the one-gluon emission shown in Fig.~\ref{fig:bJ}(b) and~\ref{fig:bJ}(c) in the threshold limit.
The first amplitude with the gluon emitted before the interaction with the current can be written in the following way:
\begin{equation}
    i\mathcal M^{(1)}_b = 
    \overline{u}(p') \Gamma \frac{i(\slashed p - \slashed k+m_Q)}{(p-k)^2-m_Q^2} (-i g_s \gamma^\mu ) u(p) \epsilon^\star_\mu
    \simeq 
    \overline{u}(p') \Gamma u(p) 
    \Bigg[
    -g_s \frac{ p \cdot \epsilon^\star}{ p \cdot k}
    \Bigg] + O(|y|/m_Q^2)
\end{equation}
where $\epsilon_\mu(k)$ and $k$ are the polarisation vector and the momentum of the gluon, respectively. 
We expanded the propagator in the limit $|k| \sim y/m_Q \ll m_Q$. 
What we obtain is the usual Eikonal factorisation.
In the second diagram \ref{fig:bJ}(c) the heavy quark propagator is
\begin{equation}
    \frac{1}{(p+q)^2-m_Q^2} =
    \frac{1}{s-m_Q^2} = -\frac{1}{y},
\end{equation}
so the amplitude factorises as follows:
\begin{equation}
    i\mathcal M^{(1)}_c = 
    \overline{u}(p')  (-i g_s \gamma^\mu ) \frac{i(\slashed p+q + m_Q)}{(p+q)^2-m_Q^2}  u(p) \epsilon^\star_\mu
    \simeq 
    \overline{u}(p') \Gamma u(p) 
    \Bigg[
    -g_s \frac{ 2 P \cdot \epsilon^\star}{ y }
    \Bigg] + O(|y|/m_Q^2),
\end{equation}
with $P = p+q$.
Around threshold, the second diagram does not reduce to the usual Eikonal term. 
Overall, around the one-particle threshold the amplitude factorises as follows:
\begin{equation}
    i\mathcal M^{(1)} = i\mathcal M^{(1)}_b+i\mathcal M^{(1)}_c \simeq
    -g_s \Bigg[
    \frac{ P \cdot \epsilon^\star}{ y }
    +\frac{ p \cdot \epsilon^\star}{ p \cdot k}
    \Bigg] 
    \, \overline{u}(p') \Gamma u(p) .
\end{equation}
Evaluating the squared amplitude, summing over the polarisation states and integrating w.r.t.\ the gluon phase-space we obtain
\begin{multline}
    W^{(1)}_\mathrm{real} =|\mathcal M^{(0)}|^2 \\
    \times (4 \pi \alpha_s) 
    \int \frac{d^3 k}{(2\pi)^3 2k_0}
    \frac{d^3 p'}{(2\pi)^3 2p'_0}
    \Bigg[
    \frac{4(m^2_Q-y)}{y^2} + \frac{m^2_Q}{(p\cdot k)^2}
    +\frac{4m^2_Q-2y-2q^2}{y (p\cdot k)}
    \Bigg]
    (2\pi)^4 
    \delta^4(P-k-p').
\end{multline}
Phase-space integration is most easily computed by expanding the integrand in the small velocity limit $|\vec v| \ll 1$ (note that before we used only the threshold limit $|y| \ll m_Q^2$).
We obtain
\begin{equation}
    W^{(1)}_\mathrm{real} = \frac{\pi}{m_Q}|\mathcal M^{(0)}|^2
    \times \frac{4 C_F }{3} \frac{\alpha_s}{ \pi } \frac{q^2}{y} \theta(-y)
    +O(y^0,(q^2)^2).
\end{equation}
We can write
\begin{equation}
    W^{(1)} = H \, U^{(1)}(\omega,\vec v) \quad \text{ with } \quad
    U^{(1)}(\omega,\vec v) =
    \frac{4 C_F }{3} \frac{\alpha_s}{ \pi } \frac{q^2}{y} \theta(-y)
    +O\left(y^0,(q^2)^2\right).
\end{equation}
This shows why the HQET parameter definitions are independent on the choice of the current:
$H$ explicitly cancel in the SV sum rules in Eq.~\eqref{eqn:SV}, leaving the infra-red universal function $U^{(1)}(\omega,\vec v) $ to control the heavy-quark mass definition.
$U^{(1)}(\omega,\vec v) $ yields the well-known perturbative version of the HQET parameters to $O(\alpha_s)$~\cite{Bigi:1996si}:
\begin{align}
    [\overline \Lambda(\mu)]_\mathrm{pert} &= 
    \lim_{\vec v\to 0} \frac{2}{\vec v^2}
    \int_0^\mu \mathrm{d}\omega\, \omega \, U^{(1)}(\omega,\vec v) = 
    \frac{4 C_F }{3} \frac{\alpha_s}{ \pi }  \mu, \notag \\
    [\mu_\pi^2]_\mathrm{pert} &= 
    \lim_{\vec v\to 0} \frac{3}{\vec v^2}
    \int_0^\mu \mathrm{d}\omega\, \omega^2 \, U^{(1)}(\omega,\vec v) = 
    C_F  \frac{\alpha_s}{ \pi }  \mu^2.
\end{align}
The mass relation to first order in $\alpha_s$ is therefore
\begin{equation}
    m_Q^\mathrm{OS} = 
    m_Q^\mathrm{kin}
    +\frac{\alpha_s}{\pi} C_F
    \left(
    \frac{4}{3} \frac{\mu}{m_Q^\mathrm{kin}}
    +\frac{1}{2}\frac{\mu^2}{(m_Q^\mathrm{kin})^2}
    + O (\mu^3)
    \right)
    +O(\alpha_s^2).
\end{equation}

\section{Details of the calculation to second and third order}
Let us now summarise the strategy of Ref.~\cite{Fael:2020njb,Fael:2020iea} for higher-order calculations. 
We generated one-,two- and three-loop four-point Feynman diagrams with \texttt{qgraf}~\cite{Nogueira:1991ex} and  used \texttt{FORM}~\cite{Ruijl:2017dtg} for algebraic manipulations. 
Afterwards, we  expanded  all  loop  momenta  according  to  the  rules  of asymptotic expansion which leads to a decomposition of each  integral  into  regions  in  which  the  individual  loop momenta either scale as \emph{hard} or \emph{ultra-soft}.  
In this respect, it was essential to choose properly the routing of loop momenta to reveal all {regions}. For each diagram we have cross-checked this using the program \texttt{asy}~\cite{Pak:2010pt}.  

The contributions where all loop momenta are hard can be discarded since there is no discontinuity.  
The regions with both hard and ultra-soft momenta are expected to cancel after renormalization and decoupling of the heavy quark from the running of the strong coupling constant.
Nevertheless we performed an explicit calculation of the \textit{uh} region at two loops, \textit{uuh} and \textit{uhh} regions at three loops. The explicit cancellation represented a crucial consistency check.  
The physical result for the quark mass relation is solely provided by the purely ultra-soft contributions.

After the various expansions, our initial four-point functions reduced to two-point functions.  
During this step, Feynman denominators become linearly dependent and a partial fraction decomposition was needed in order to obtain integral families with linearly independent propagators. To automate the procedure we used the program \texttt{LIMIT}~\cite{Herren:2020ccq}.
Reduction of the integrals to master integrals was performed with the programs \texttt{FIRE}~\cite{Smirnov:2019qkx} and \texttt{LiteRed}~\cite{Lee:2013mka}.

We found 1, 3 and 20 ultra-soft  master integrals at one-, two- and three-loop order which needed a novel analytic calculation.
At one and two loops they are easily expressed in terms of Euler's $\Gamma$ functions
with exact dependence on the dimensional regulator $\epsilon= (4-d)/2$. 
Also eleven of the three-loop master integrals have such convenient form.  
For the remaining eight integrals, we obtained the necessary terms in the $\epsilon$-expansion through Mellin-Barnes representations and the use of \texttt{MB} package~\cite{Czakon:2005rk,Smirnov:2009up}.  
The analytic expressions were obtained by closing the integration contour of the Mellin-Barnes integrals and summing the pole residues with \texttt{Sigma}~\cite{schneider2013simplifying}, or also by performing high-precision numerical integrations and utilising the PSLQ algorithm~\cite{MR1489971} to reconstruct the analytic form. 
For one integral where the Mellin-Barnes method was not successful, we introduced a second mass scale $x$ in one of the heavy quark propagators. 
We derived a set of differential equations~\cite{Gehrmann:1999as,Henn:2013pwa} in $x$, applied  boundary  conditions  at $x=0$ where the integrals could be analytically computed,  and then evaluated  the solution for $x= 1$ which provided the desired results.

We considered also for the first time finite charm mass effects in the mass relation for the bottom quark both at $O(\alpha_s^2)$ and $O(\alpha_s^3)$. 
All charm quark mass effects in the $n_l= 4$ flavour theory are decoupling effects, even if
the bare three-loop diagrams have a non-trivial dependence on $m_c/m_b$ which eventually cancels out after renormalization. 
Therefore, the kinetic mass for the bottom can be expressed in term of $\alpha_s^{(3)}$ without an explicit occurrence of $m_c$. 
The transition from $\alpha_s^{(3)}$ to $\alpha_s^{(4)}$  generates finite $m_c$ effects as $\log(\mu_\mathrm{dec}^2/m_c^2)$ terms, where $\mu_\mathrm{dec}$ is the scale where the charm quark is decoupled.

\section{Results} 
The explicit expressions for the relation between the kinetic and the on-shell mass is given in~\cite{Fael:2020njb,Fael:2020iea}.  To obtain the conversion formula between the $\MSbar$ and the kinetic scheme,  we replaced the on-shell mass with the $\MSbar$ mass  using the corresponding three-loop relation~\cite{Chetyrkin:1999qi,Melnikov:2000qh,Chetyrkin:1999ys,Fael:2020bgs}.
Our results are implemented in the programs \texttt{(C)RunDec}~\cite{Herren:2017osy} and \texttt{REvolver}~\cite{Hoang:2021fhn}.

As an example, we show how to convert with \texttt{RunDec} the bottom mass from the $\MSbar$ scheme to the kinetic scheme.
We use as input parameters the FLAG averaged values~\cite{FlavourLatticeAveragingGroup:2019iem} with $N_f=2+1+1$ quark flavours: 
$ \overline{m}_b(\overline{m}_b) =  4.198(12)$ GeV
and $\overline{m}_c(3 \text{ GeV})  =0.988 (7)$ GeV. 
We set the Wilsonian cutoff $\mu= \mu_\mathrm{WC} =1$ GeV and  $\alpha_s^{(5)}(M_Z)  =  0.1179$~\cite{ParticleDataGroup:2020ssz}.
In Mathematica, first we load the \texttt{RunDec} package and set the relevant parameters:
\begin{verbatim}
In[] := <<RunDec.m;
In[] := {mbMS, mub, mcMS, muc} = {4.198, 4.198, 0.988, 3};
In[] := {as4, mus, muWC} = {0.224262, mbMS, 1};
\end{verbatim}
The function that converts $\overline{m}_b(\overline{m}_b)$ to $m_b^\mathrm{kin}$ is \texttt{mMS2mKIN} which receives as input values $\overline{m}_b(\mu_b)$, $\{\overline{m}_c(\mu_c),\mu_c\}$, $\alpha_s^{(n_f)}(\mu_s)$, $\mu_s$, $\mu_\mathrm{WC}$, $n_\mathrm{loops}$ and a string either \verb!"A","B","C"! or \verb!"D"!, which selects one of the schemes for the treatment of $m_c$ effects considered in~\cite{Fael:2020njb}.
The value obtained by \texttt{RunDec} is
\begin{verbatim}
   In[] := mbK = mMS2mKIN[mbMS, {mcMS, muc}, AA*as4, mus, muWC, 3, "B"]
   In[] := mbK /. AA->1
   Out[] = 4.198 + 0.261322 AA + 0.0787718 AA^2 + 0.0268024 AA^3
   Out[] = 4.5649
\end{verbatim}
To  estimate  the  theoretical  uncertainty  associated  to  the conversion formula
we use half of the three-loop correction as the size of the unknown higher orders.  This leads to an uncertainty of about 15~MeV to $O(\alpha_s^3)$. The same approach applied to the two-loop mass relation leads to an uncertainty of about 40 MeV so the three-loop formula reduces the uncertainty by about a factor of two.
Our final prediction for the kinetic mass of the bottom quark is
\begin{equation}
    m_b^\mathrm{kin}(1 \text{ GeV}) = 
    4.565 \, (15)_\mathrm{th} (13)_\mathrm{lat} \text{ GeV}=
    4.565 \, (20) \text{ GeV},
\end{equation}
where the first uncertainty (th) comes from the conversion formula while the second (lat) is the error of $\overline{m}_b (\overline{m}_b)$ from lattice average.

\section{Conclusion}
We presented the relation between the on-shell mass and the kinetic mass, showing an explicit calculation of the first-order correction.
We also summarised the method which we employed to determine the conversion formula up to three-loop order. This was achieved by a suitable threshold expansion of the relevant structure function entering the SV sum rules.
Our strategy is in principle extendable to four loops, if such precision will ever become necessary in the future.  
Our results can be most easily utilised via the \texttt{RunDec} or \texttt{REvolver} implementations.
The  new  correction terms at three loops reduce the uncertainty due to scheme conversion by about a factor two. 
Our results have been crucial to reduce the uncertainty in the inclusive $|V_{cb}|$ determination~\cite{Bordone:2021oof} and they will be also necessary in future theoretical updates of $|V_{cb}|$ after incorporating new Belle II data.

\section*{Acknowledgements}
I thank K.\ Sch\"onwald and M.\ Steinhauser for the nice and fruitful collaboration. I thank also P.\ Gambino and T.\ Mannel for discussion and correspondence. This  research  was  supported  by  the  Deutsche  Forschungsgemeinschaft  (DFG,  German  ResearchFoundation) under grant 396021762 -- TRR 257 ``Particle Physics Phenomenology after the Higgs Discovery''.

\bibliography{BIB.bib}

\nolinenumbers

\end{document}